\documentclass[conference]{IEEEtran}
\IEEEoverridecommandlockouts
\usepackage{cite}
\usepackage{amsmath,amssymb,amsfonts}
\usepackage{epsfig}
\usepackage{theorem}
\usepackage{graphicx}
\usepackage{textcomp}
\usepackage{xcolor}
\def\BibTeX{{\rm B\kern-.05em{\sc i\kern-.025em b}\kern-.08em
    T\kern-.1667em\lower.7ex\hbox{E}\kern-.125emX}}

\begin{document}

\title{Adaptive Transmit Waveform Design
\thanks{This research was funded by the Office of Naval Research and NUWC's internal investment program.}
}

\author{\IEEEauthorblockN{David A. Hague}
\IEEEauthorblockA{\textit{Sensors and Sonar Systems Department} \\
\textit{Naval Undersea Warfare Center}\\
Newport, RI USA\\
david.a.hague@navy.mil}
}
\maketitle

%\section{Introduction}
Recent research efforts in the Anti-Submarine Warfare (ASW) community have focused on developing sonar systems that adapt to their acoustic environment, referred to as “cognitive” sonars.  Cognitive active sonar systems utilize principles of the perception action cycle of cognition to leverage information gathered from earlier sensing interactions with the underwater acoustic environment \cite{Nelson}.  This in turn informs the selection of system parameters to optimize target detection, classification, localization, and tracking performance in that acoustic environment.  Of the many system parameters such a cognitive sonar system could potentially adapt, the acoustic signal transmitted into the medium, also known as the transmit waveform, has a profound impact on system performance.   Many of the physical characteristics of the acoustic environment are contained in the return echo signal that is composed of amplitude scaled (target strength), time-delayed (target range) and Doppler shifted (target range-rate) echoes of the transmit waveform.  The active sonar system then processes these echo signals typically with a bank of Matched Filters (MF) tuned to an array of potential target Doppler values.  The output of such a receiver yields a range-Doppler map of the target scene \cite{Ricker}.  

Of the many transmit waveform parameters cognitive sonar systems could adapt including pulse length, transmit source level, and the operational band of frequencies, the type of phase or frequency modulation employed by the transmit waveform also fundamentally influences the type and quality of the information inferred about the target scene \cite{Blunt}.  Additionally, the waveform should facilitate efficient transmission on piezoelectric transducers and their driving electronics.  It should possess a constant amplitude to minimize the distortion resulting from saturated power amplifiers, which drive the projector transducer.  Constant amplitude also maximizes the energy transmitted into the medium for given duration and peak transmit power limits.  Waveforms should also be spectrally compact; the vast majority of the waveform’s energy should be concentrated densely in the operational band with very little energy elsewhere.  This reduces mutual interference between systems operating in adjacent frequency bands and any distortion in the resulting transmitted acoustic signal from the frequency dependent filtering of the projector transducer and its driving electronics.   

There has been a wealth of research on waveform design dating back several decades.  Continuous Wave (CW) waveforms, perhaps the simplest of all sonar waveforms, possesses a constant frequency throughout their duration and achieves high Doppler resolution but poor range resolution.  The Linear Frequency Modulated (LFM) waveform, the first of the class of pulse compression waveforms, possesses both large bandwidth and long duration resulting in a large Time-Bandwidth Product (TBP).  The large bandwidth provides improved target resolution and the long duration provides the necessary transmit energy for good target detection performance in noise-limited conditions.  In the 1960’s, the Hyperbolic FM (HFM) was introduced as a large TBP \cite{Jan} counterpart to the LFM that was robust to Doppler mismatch and has found use in broadband active sonar applications \cite{Ricker}.  Also in the 1960’s, John Costas developed a family of frequency-shift keying (FSK) waveforms that jointly resolved target range and Doppler in a single waveform \cite{Ricker, CostasI}.  An FSK waveform is composed of equal length CW sub-pulses, known as chips, which are contiguous in time.  Each chip possesses a different center frequency according to a particular “firing” code.  Costas defined the basic necessary properties for these firing codes that now bear his name \cite{CostasII}.  Comb waveforms, whose spectral shape resembles the Dirac comb function, distinguish moving targets in stationary reverberation \cite{Comb}.  All of these waveform types address a particular active sonar design problem and are rather diverse in their characteristics.  A cognitive sonar system may very well determine that the optimal waveform for a given scenario is one of the aforementioned waveform types.

However, fully leveraging the adaptive capability of a cognitive sonar system requires a waveform model that facilitates adaptation according to a set of system defined goals and design metrics.  The vast majority of adaptive transmit waveform design research has focused on Poly-Phase Coded (PPC) waveforms from the radar literature \cite{Blunt}. A PPC waveform is composed of a train of equal duration CW chips contiguous in time all sharing a common center frequency.  The phase of each chip is then assigned different values in a manner that generates a waveform with the desired characteristics.  There exists a nearly endless combination of realizable poly-phase codes making PPC waveforms far more versatile than their FM and FSK counterparts.  While FM and FSK waveforms are limited in their versatility, they are readily implementable on practical systems due to their constant amplitude and spectral compactness properties.  PPC waveforms on the other hand, suffer from substantial spectral extent due to the nearly instantaneous phase transitions between chips.  This has motivated the development of Continuous Phase Modulation (CPM) techniques to improve upon their spectral characteristics by smoothing the phase transitions between chips \cite{Blunt}.  This phase smoothing essentially transforms PPC waveforms into constant amplitude, spectrally efficient parameterized FM waveforms.  However, this smoothing also introduces perturbations to the waveform’s characteristics, which then requires re-optimization of the original PPC waveform’s phase-code.   

Inspired by these CPM efforts, the author developed a spectrally compact adaptive FM waveform model using Multi-Tone Sinusoidal Frequency Modulation (MTSFM) \cite{Hague_AES}. The MTSFM waveform’s frequency and phase modulation functions are composed of a finite set of weighted sinusoidal harmonics.  The weights for each harmonic are utilized as a discrete set of design coefficients.  Adjusting these coefficients results in constant amplitude, spectrally compact FM waveforms with unique characteristics.  Figure 1 shows an example MTSFM waveform.  The spectrogram of the waveform, which conveys its time-frequency structure, possesses smooth oscillatory characteristics unlike a PPC waveform.  As a result, the waveform’s spectral energy is densely concentrated in the swept bandwidth B with very little energy residing outside of that band.  This waveform is optimized to possess low sidelobes in a specified region in range of its MF output (denoted by the red dashed lines) which allows for distinguishing weak echoes in the presence of a much stronger one.  These sidelobes are substantially lower than the MF response of a Costas waveform with the same TBP (shown in blue).  

The motivation for this design example is primarily to demonstrate how the MTSFM’s design parameters can be modified to finely tune the resulting waveform’s characteristics.  However, this design may also find use in certain scenarios.  Real-world target returns are usually composed of multiple echoes from the acoustic highlights of the target, which can vary greatly in strength.  Distinguishing these echoes infers details about the physical makeup of that target.  Figure 2 shows the advantages the example MTSFM design has in distinguishing a collection of closely spaced echoes over the Costas waveform from Figure 1.  The MF response from the Costas waveform (top panel) only picks out the stronger echoes and masks three of the weaker ones.  The MTSFM however easily distinguishes each echo due to the much lower range sidelobes.  

The design example shown in Figures 1 \& 2 is only one of many waveform design problems where the MTSFM waveform model is applicable.  The MTSFM model facilitates adjustable Doppler tolerance.  In fact, it can smoothly transition from being ideally Doppler sensitive like a FSK or CW waveform, to being ideally Doppler tolerant like an LFM or HFM waveform \cite{Hague_EOA}.  This ability offers the designer a tradeoff between target Doppler resolution and receiver complexity (i.e, the number of required MFs to process the target scene).  Additionally, other characteristics of the waveform such as range and Doppler sidelobes can also be further refined while maintaining the desired Doppler tolerance/sensitivity.  The MTSFM can also synthesize Comb waveforms for distinguishing moving targets in stationary reverberation \cite{Hague_UASP_2019}.  Comb waveform design typically focuses on ensuring strong reverberation suppression at specific target Doppler values while also possessing a constant amplitude and reasonably low range sidelobes \cite{Comb}.  Generally, a waveform cannot achieve all three of these considerations simultaneously but can tradeoff between them.  Geometric comb waveforms \cite{Cox} have to date represented one of the best tradeoff designs between these three considerations.  The MTSFM model also can smoothly trade-off between these design considerations and while not superior to the Geometric comb waveform, is at least competitive in its design characteristics.  Lastly, another application of the MTSFM focuses on designing not just one waveform, but families of MTSFM waveforms that occupy a common band of frequencies and also possess low cross-interference properties with one another \cite{HagueFam}.  Such waveforms may be applicable for use in multi-static active sonar systems where reducing the mutual interference between each waveform is paramount.  The MTSFM closely approaches established performance bounds of such waveform families in a manner similar to that of FSK and PPC waveforms geared towards the same application.  The adaptability of the MTSFM waveform model allows it to possess a wide variety of performance characteristics that in the past has required a diverse set of waveform designs to achieve.

The MTSFM is an adaptive waveform that synthesizes constant amplitude, spectrally compact waveforms that can possess a wide variety of desirable properties.  The adaptability of the MTSFM combined with its transmitter friendly properties make it an attractive waveform type for a variety of active sonar applications.  The intent of employing the MTSFM waveform model is not to outright replace the many waveform types used by current sonar systems.  Rather, the intention is to provide a cognitive sonar system the ability to generate a complementary set of finely tuned waveforms for the novel scenarios and environments that it may encounter.  In this sense, the MTSFM waveform may very well be an enabler for cognitive active sonar systems.  


\begin{figure*}[ht]
\centering
\includegraphics[width=1.0\textwidth]{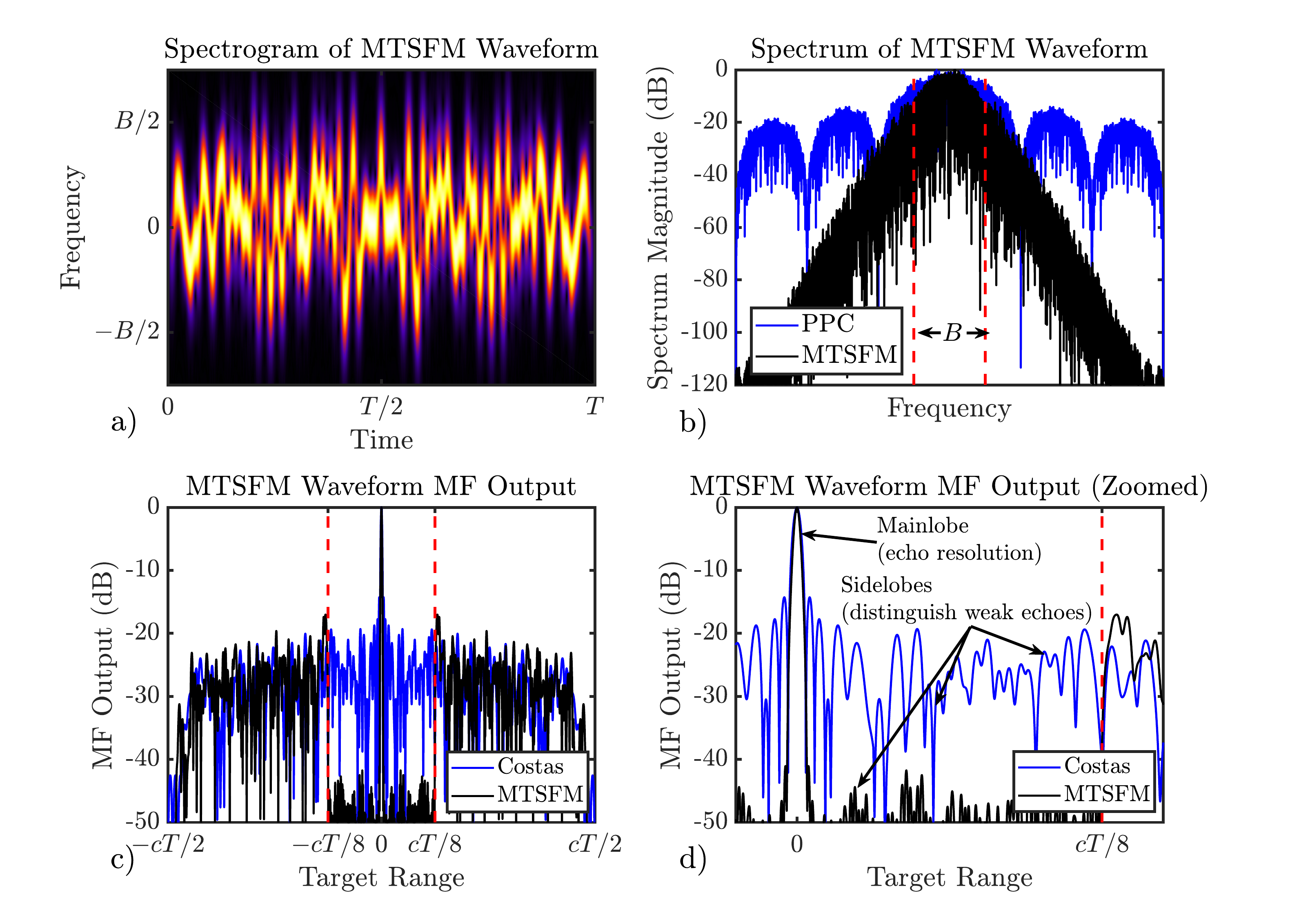}
\caption{Spectrogram (a), spectrum (b), and MF response (c \& d) of an example MTSFM waveform with a TBP of 256.  Also shown in (b) is the spectrum of an equivalent bandwidth PPC waveform and in (c \& d) the range response of a Costas waveform with equivalent TBP.  The waveform’s range response is optimized to reduce sidelobe levels in the region denoted by the red dashed lines.  This design allows for distinguishing several closely echoes with varying strengths while possessing a spectrum whose energy is densely concentrated in the waveform’s swept bandwidth $B$.}
\label{fig:MTSFM_1}
\end{figure*}

\begin{figure*}[ht]
\centering
\includegraphics[width=1.0\textwidth]{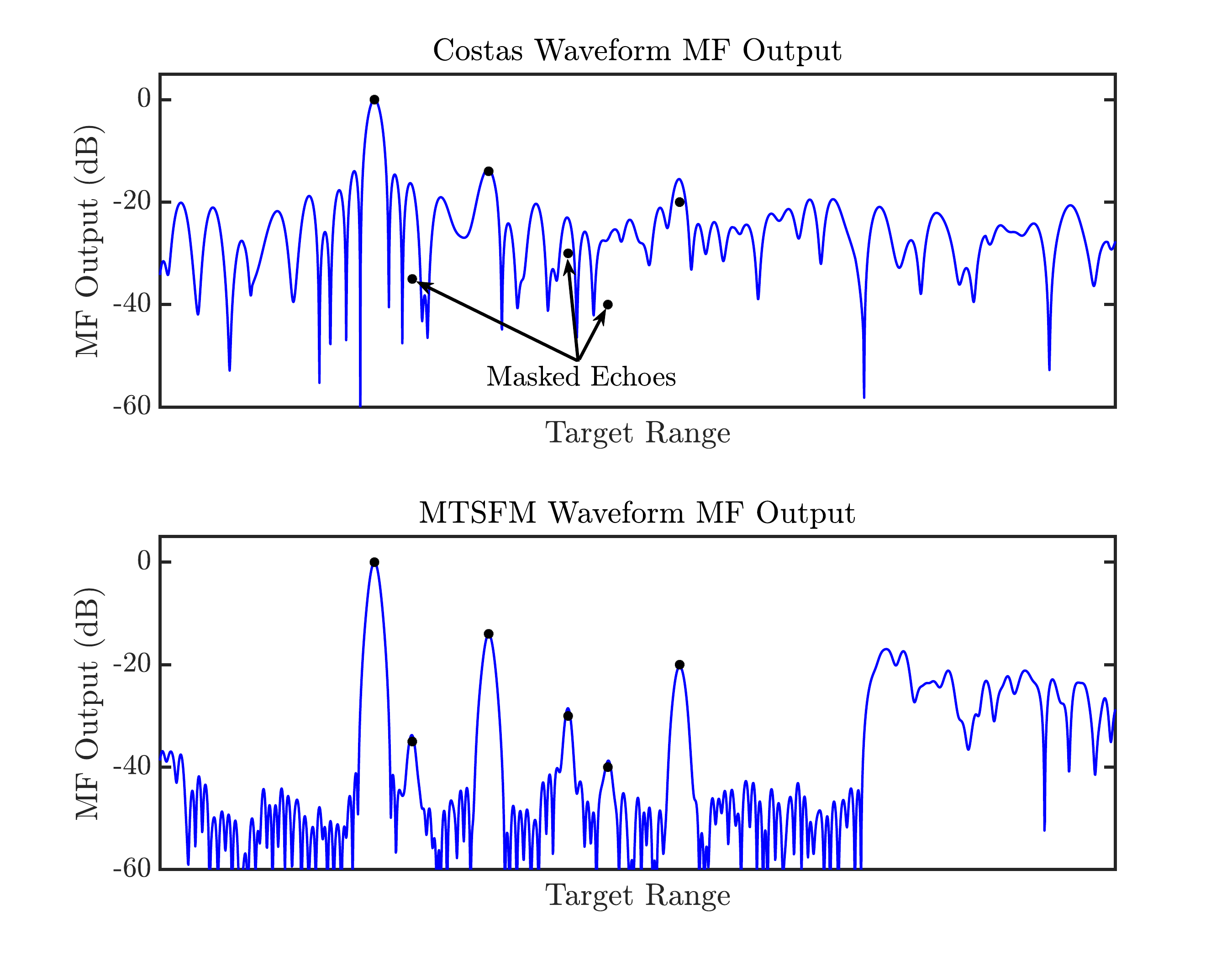}
\caption{MF output for the Costas and MTSFM waveforms from Figure 1 for a collection of closely spaced echoes (denoted by the black dots) with a 40 dB variation in echo strength.  While the Costas waveform’s sidelobe levels are too high to distinguish the weaker echoes, the MTSFM’s suppressed sidelobes allow for distinguishing even the weakest echo in the presence of the strongest one}
\label{fig:MTSFM_2}
\end{figure*}

\end{document}